\documentclass[Conference,a4paper]{IEEEtran}

\IEEEoverridecommandlockouts

\usepackage{color}
\usepackage{graphicx}
\usepackage{amsmath}
\usepackage{amssymb}
\usepackage{algorithm}
\usepackage{algorithmic}
\usepackage{amsmath}
\usepackage{multirow}
\usepackage{booktabs}
\usepackage{array}
\usepackage{amsthm}
\usepackage{stfloats}
\usepackage{caption}
\usepackage{subfigure}
\usepackage{bm}
\usepackage{booktabs}
\usepackage{setspace}
\usepackage{gensymb}
\usepackage{url}

\usepackage{flushend}
\usepackage{lastpage}

\newcommand{\be}{\begin{equation}}
\newcommand{\ee}{\end{equation}}
\newcommand{\bea}{\begin{eqnarray}}
\newcommand{\eea}{\end{eqnarray}}
\newcommand{\ba}{\begin{array}}
\newcommand{\ea}{\end{array}}

\newcommand{\non}{\nonumber}

\allowdisplaybreaks[4]

\title{CRB Optimization using a Parametric Scattering Model for Extended Targets in ISAC Systems}
\author{\IEEEauthorblockN{Rang Liu$^{\dag}$, A. Lee Swindlehurst$^{\dag}$, and Ming Li$^{\ddag}$ }\\
\IEEEauthorblockA{$^{\dag}$University of California, Irvine, CA 92697, USA\\ 
\IEEEauthorblockA{$^{\ddag}$Dalian University of Technology, Dalian, Liaoning 116024, China \\ E-mail: \texttt{\{rangl2, swindle\}@uci.edu; mli@dlut.edu.cn} } }}

\begin{document}

\maketitle

\pagestyle{empty}
\thispagestyle{empty}

\begin{abstract}
This paper presents a novel parametric scattering model (PSM) for sensing extended targets in integrated sensing and communication (ISAC) systems. The PSM addresses the limitations of traditional models by efficiently capturing the target’s angular characteristics through a compact set of key parameters, including the central angle and angular spread, enabling efficient optimization. Based on the PSM, we first derive the Cram{\'e}r-Rao Bound (CRB) for parameter estimation and then propose a  beamforming design algorithm to minimize the CRB while meeting both communication signal-to-interference-plus-noise ratio (SINR) and power constraints. By integrating the PSM into the beamforming optimization process, the proposed framework achieves superior CRB performance while balancing the tradeoff between sensing accuracy and communication quality. Simulation results demonstrate that the PSM-based approach consistently outperforms traditional unstructured and discrete scattering models, particularly in resource-limited scenarios, highlighting its practical applicability and scalability.
\end{abstract}

\section{Introduction}\label{sec:introduction}

Integrated sensing and communication (ISAC) empowers wireless infrastructure to simultaneously perform sensing and communication, establishing it as a key enabling technology for sixth-generation (6G) systems \cite{LiuFan JSAC 2022}-\cite{AK2024}. In ISAC systems, the design of dual-functional waveforms is essential to balance radar and communication performance. Recent research has emphasized beamforming design to exploit the available spatial degrees of freedom (DoFs), exploring advanced techniques across various scenarios.

For sensing in ISAC systems, most work assumes point-like targets to simplify the mathematical formulations and associated beamforming designs. However, modeling the target as a single reflection point is inadequate for many applications such as those encountered in automotive radar, where reflections from objects such as vehicles and roadside infrastructure can cover a relatively wide angular range. Moreover, with modern large antenna arrays offering high angular resolution, sensed targets often span multiple angular bins. Therefore, it is necessary to incorporate extended target models into beamforming designs for ISAC systems.

There has been limited prior work that has investigated beamforming designs for ISAC systems with an extended target. In one approach, each element of the unstructured target response matrix is estimated, and the corresponding Cram{\'e}r-Rao bound (CRB) is formulated to guide beamforming optimization \cite{Liu TSP 2022}-\cite{Song TWC 2024}. While these approaches can extract angular information about the target from the channel matrix, the unstructured channel model (UCM) used in the CRB optimization fails to leverage the target's inherent angular structure, resulting in suboptimal performance. Additionally, estimating a high-dimensional response matrix introduces significant computational complexity.
In contrast, the discrete scattering model (DSM) provides a more refined representation by capturing the angular information of individual scatterers along the target, offering an alternative framework for more detailed target modeling \cite{Karbasi TSP 2015}-\cite{Du TWC 2023}. However, DSM  emphasizes estimating each scattering point on the target while ignoring their inherent physical connection across the same target, leading to over-parameterization and also a higher-dimensional optimization. Therefore, beamforming designs based on UCM and DSM will consume more of the available spatial DoFs, limiting their ability to also guarantee the communication performance required for ISAC operation.

Motivated by the above, we introduce a novel parametric scattering model (PSM) to efficiently describe extended targets in ISAC applications. The PSM captures key angular characteristics such as the central angle and angular spread of the target, enabling a more efficient optimization without compromising estimation accuracy. Based on the PSM, we formulate the CRB for estimating the reduced set of target parameters. Then, the transmit beamforming is optimized to minimize the CRB under communication signal-to-interference-plus-noise ratio (SINR) and transmit power constraints. Simulation results demonstrate that the PSM-based design achieves a significantly lower CRB compared to the designs based on UCM and DSM, especially when the antenna resources are limited. These results highlight the superiority of the proposed PSM and associated beamforming design in effectively utilizing the angular structure of extended targets, showcasing their potential for practical implementation in real-world ISAC applications.

\section{System Model}
We consider a monostatic ISAC system in which a base station (BS) is equipped with two arrays, one designated for transmission and one for reception. The transmit array comprising $N_\text{t}$ elements emits dual-functional signals to simultaneously communicate with $K$ single-antenna users and illuminate a target. Meanwhile, the receive array consisting of $N_\text{r}$ elements captures the echo signals for target parameter estimation.  In this paper, we consider a scenario with a target that extends across a range of angles, such as a vehicle or other large object that. The extended target reflects signals from multiple points across its surfaces, producing a contiguous set of echoes. Since the extended target spans multiple angular bins in the spatial domain, accurately modeling their reflection characteristics is crucial for effective parameter estimation and beamforming optimization.
 
The BS transmits a superposition of the precoded communication symbols $\mathbf{s}_\text{c}\in\mathbb{C}^K$ and sensing signals $\mathbf{s}_\text{r}\in\mathbb{C}^{N_\text{t}}$, resulting in the following dual-functional transmitted signal:
\be
\mathbf{x} = \mathbf{W}_\text{c}\mathbf{s}_\text{c} + \mathbf{W}_\text{r}\mathbf{s}_\text{r} = \mathbf{W}\mathbf{s},
\ee
where $\mathbf{W}_\text{c}\in\mathbb{C}^{N_\text{t}\times K}$ and $\mathbf{W}_\text{r}\in\mathbb{C}^{N_\text{t}\times N_\text{t}}$ represent the beamforming matrices for communication and sensing, respectively.  We define the overall beamforming matrix as $\mathbf{W}\triangleq[\mathbf{W}_\text{c}~\mathbf{W}_\text{r}]$ and the symbol vector as $\mathbf{s}\triangleq[\mathbf{s}_\text{c}^T~\mathbf{s}_\text{r}^T]^T$, and we assume that 
$\mathbb{E}\{\mathbf{ss}^H\} = \mathbf{I}_{N_\text{t}+K}$. The signal received at the $k$-th user is 
\be
y_k = \mathbf{h}_k^T\mathbf{Ws} + n_k,
\ee
where $\mathbf{h}_k\in\mathbb{C}^{N_\text{t}}$ denotes the channel between the BS and the $k$-th user and $n_k\sim\mathcal{CN}(0,\sigma_k^2)$ is  additive white Gaussian noise (AWGN). The SINR at the $k$-th user is calculated as 
\be
\text{SINR}_k = \frac{|\mathbf{h}_k^T\mathbf{w}_{k}|^2}
{\sum_{j\neq k}^{K+N_\text{t}}|\mathbf{h}_k^T\mathbf{w}_j|^2+\sigma_k^2}.
\ee

The radar returns collected during $L$ symbol slots at the BS receive array can be expressed as 
\be
\mathbf{Y}_\text{r} = \mathbf
{GWS}+ \mathbf{Z},
\ee
where $\mathbf{G}\in\mathbb{C}^{N_\text{r}\times N_\text{t}}$ represents the target response matrix, $\mathbf{S}\in\mathbb{C}^{(N_\text{t}+K)\times L}$ contains the communication and radar signals over the $L$ symbol slots, and $\mathbf{Z}\in\mathbb{C}^{N_\text{r}\times L}\sim\mathcal{CN}(\mathbf{0},\sigma_\text{r}^2\mathbf{I}_{N_\text{r}})$ is AWGN. Under UCM, the radar signature of the target is described by the unstructured matrix $\mathbf{G}$. While the target parameters enter the model in a linear way, the number of parameters $N_r N_t$ that must be estimated can be very large, and the model ignores information that is available about the transmit and receive array geometries and the physical dimensions of the target. In fact, the UCM could be used to describe multiple point-like targets that are distributed arbitrarily in space with no relationship to each other, as would be the case for an extended target.

Alternatively, the parametric DSM approach is often used in the literature to describe extended targets, where the target is represented as a collection of point-like scatterers, each contributing independently to the reflected signal. Under this model, the target response matrix is given by
\be\label{eq:G}
\mathbf{G} = \sum_{t=1}^T\alpha_t\mathbf{b}(\theta_t)\mathbf{a}^T(\theta_t),
\ee
where $T$ is the number of scatterers, $\{\theta_1,\theta_2,\ldots,\theta_T\}$ denote the angles of the scatterers, $\alpha_t$ is the complex channel gain or radar cross section (RCS) of the $t$-th scatterer, and $\mathbf{b}(\theta_t)\in\mathbb{C}^{N_\text{r}}$ and $\mathbf{a}(\theta_t)\in\mathbb{C}^{N_\text{t}}$ are the receive and transmit array response vectors, respectively. 
Under DSM, the extended target parameters are $\{\theta_t,~\alpha_t,~\forall t\}$. While this model incorporates information about the transmit and receive array responses, no information is exploited about the fact that the scatterers are part of a single extended target for which the angles $\theta_t$ are relatively closely spaced. For large $T$, this model overparameterizes the problem, and can also lead to high computational complexity and inaccurate angle estimates. 

Instead, as described in the next section, we focus on estimating a small set of key angle parameters that effectively capture the target’s essential characteristics while ensuring improved accuracy and computational efficiency. 

\section{Parametric Scattering Model and Problem Formulation}
In this section, we introduce a parametric scattering model (PSM) to effectively characterize extended targets and facilitate CRB-oriented beamforming optimization. Rather than estimating the angles for many individual discrete scatterers, PSM employs a higher-level description involving the center angle $\theta_0$ of the target and its angular spread  $\Delta$ to provide a concise yet accurate representation of the extended target. While the number of angle-related parameters is significantly reduced, PSM still retains the ability to model how the RCS varies across the target. This more detailed RCS modeling is manageable since the RCS terms are linear parameters and can be estimated in closed-form.

Like DSM, for PSM we assume that the target is composed of multiple distributed scatterers whose reflections contribute to the overall target response. Unlike DSM, the angles $\{\theta_1,\theta_2,\ldots,\theta_T\}$ of the target scatterers are assumed to be uniformly spaced in the interval $[\theta_0-\Delta/2,\theta_0+\Delta/2]$. Thus, the angle of the $t$-th scatter can be expressed as a function of $\theta_0$ and $\Delta$ as 
\be\label{eq:thetat}
\theta_t = \theta_0+\frac{2t-T-1}{2(T-1)}\Delta.
\ee
While the angles of the individual scatterers are specified in terms of the target's central angle and angular spread, the RCS parameters associated with the scatterers are retained. Thus, the parameters to be estimated using PSM are defined by $\bm{\xi} \triangleq [\theta_0,~ \Delta,~ \Re\{\bm{\alpha}\},~ \Im\{\bm{\alpha}\}]^T$, where $\bm{\alpha} \triangleq [\alpha_1,\alpha_2,\ldots,\alpha_T]^T$ represents the RCS vector. Thus, for PSM the target response matrix $\mathbf{G}$ is reformulated as a function of $\bm{\xi}$.

Based on the above descriptions of PSM, we derive the CRB for estimating $\bm{\xi}$ in order to develop the subsequent beamforming design. We first vectorize the received signal as 
\be\label{eq:vector signal}
\mathbf{y}_\text{r}\triangleq\text{vec}\{\mathbf{Y}_\text{r}\} = \bm{\eta} + \mathbf{z} = 
\text{vec}\{\mathbf{GWS}\} + \mathbf{z},
\ee
where we define $\bm{\eta}\triangleq\text{vec}\{\mathbf{GWS}\}$ and $\mathbf{z}\triangleq\text{vec}\{\mathbf{Z}\}\sim\mathcal{CN}(\mathbf{0},\sigma_\text{r}^2\mathbf{I}_{N_\text{r}L})$. Then, we write the $(i,j)$-th element of the Fisher information matrix $\mathbf{F}\in\mathbb{R}^{(2T+2)\times(2T+2)}$ as
\be
\mathbf{F}(i,j) = \frac{2}{\sigma_\text{r}^2}\Re\Big\{\frac{\partial^H\bm{\eta}}{\partial\xi_i}\frac{\partial\bm{\eta}}{\partial\xi_j}\Big\}.
\ee
According to the signal model in (\ref{eq:vector signal}) and the target response matrix in (\ref{eq:G}), the partial derivatives can be calculated by 
\begin{subequations}\begin{align}
\frac{\partial\bm{\eta}}{\partial\theta_0} &= \text{vec}\{\mathbf{G}_{\theta_0}\mathbf{WS}\},\\
\frac{\partial\bm{\eta}}{\partial\Delta} &= \text{vec}\{\mathbf{G}_\Delta\mathbf{WS}\},\\
\frac{\partial\bm{\eta}}{\partial \Re\{\alpha_t\}} &= \text{vec}\{\mathbf{G}_t\mathbf{WS}\},\\
\frac{\partial\bm{\eta}}{\partial \Im\{\alpha_t\}} &= \jmath\text{vec}\{\mathbf{G}_t\mathbf{WS}\},
\end{align}\end{subequations}
where we define 
\begin{subequations}\label{eq:G derivatives}\begin{align}
\mathbf{G}_{\theta_0}& \triangleq\frac{\partial \mathbf{G}}{\partial \theta_0} =  \sum_{t=1}^T\alpha_t[\dot{\mathbf{b}}(\theta_t)\mathbf{a}^T(\theta_t) + \mathbf{b}(\theta_t)\dot{\mathbf{a}}^T(\theta_t)], \\
\mathbf{G}_\Delta& \triangleq\frac{\partial \mathbf{G}}{\partial \Delta} \non\\
&= \sum_{t=1}^T\!\alpha_t\frac{2t\!-\!T\!-\!1}{2(T\!-\!1)}[\dot{\mathbf{b}}(\theta_t)\mathbf{a}^T(\theta_t) \!+\! \mathbf{b}(\theta_t)\dot{\mathbf{a}}^T(\theta_t)],\!\\
\mathbf{G}_t&\triangleq\frac{\partial \mathbf{G}}{\partial \Re\{\alpha_t\}} = \mathbf{b}(\theta_t)\mathbf{a}^T(\theta_t),
\end{align}\end{subequations}
$\dot{\mathbf{b}}(\theta_t) = \partial \mathbf{b}(\theta_t)/\partial\theta_0 = \partial \mathbf{b}(\theta_t)/\partial\theta_t$, and similarly for $\dot{\mathbf{a}}(\theta_t)$.
The elements of $\mathbf{F}$ are then given by
\begin{subequations}\label{eq:F elements}\begin{align}
    \mathbf{F}(1,1)&= \frac{2L}{\sigma^2_\text{r}}\Re\{\text{Tr}\{\mathbf{G}_{\theta_0}\mathbf{R}_\text{w}\mathbf{G}_{\theta_0}^H\}\},\\
    \mathbf{F}(1,2)&= \frac{2L}{\sigma^2_\text{r}}\Re\{\text{Tr}\{   \mathbf{G}_\Delta\mathbf{R}_\text{w}\mathbf{G}_{\theta_0}^H\}\},\\
    \mathbf{F}(2,2)&= \frac{2L}{\sigma^2_\text{r}}\Re\{\text{Tr}\{   \mathbf{G}_\Delta\mathbf{R}_\text{w}\mathbf{G}_\Delta^H\}\},\\
    \mathbf{F}(1,i\!+\!2)&= \frac{2L}{\sigma^2_\text{r}}\Re\{\text{Tr}\{   \mathbf{G}_i\mathbf{R}_\text{w}\mathbf{G}_{\theta_0}^H\}\},\\ 
    \mathbf{F}(1,T\!+\!i\!+\!2)&= 2L\Re\{\jmath\text{Tr}\{   \mathbf{G}_i\mathbf{R}_\text{w}\mathbf{G}_{\theta_0}^H\}\},\\
        \mathbf{F}(2,i\!+\!2)&= \frac{2L}{\sigma^2_\text{r}}\Re\{\text{Tr}\{   \mathbf{G}_i\mathbf{R}_\text{w}\mathbf{G}_\Delta^H\}\},\\ 
        \mathbf{F}(2,T\!+\!i\!+\!2)&= \frac{2L}{\sigma^2_\text{r}}\Re\{\jmath\text{Tr}\{\mathbf{G}_i\mathbf{R}_\text{w}\mathbf{G}_\Delta^H\}\},\\ 
    \mathbf{F}(i\!+\!2,j\!+\!2)&= \frac{2L}{\sigma^2_\text{r}}\Re\{\text{Tr}\{   \mathbf{G}_j\mathbf{R}_\text{w}\mathbf{G}_i^H\}\},\\    
        \mathbf{F}(i\!+\!T\!+\!2,j\!+\!2)&= \frac{2L}{\sigma^2_\text{r}}\Re\{-\jmath\text{Tr}\{\mathbf{G}_j\mathbf{R}_\text{w}\mathbf{G}_i^H\}\},\\ 
      \mathbf{F}(i\!+\!2,j\!+\!T\!+\!2)&= \frac{2L}{\sigma^2_\text{r}}\Re\{\jmath\text{Tr}\{\mathbf{G}_j\mathbf{R}_\text{w}\mathbf{G}_i^H\}\},\\ 
       \mathbf{F}(i\!+\!T\!+\!2,j\!+\!T\!+\!2)&= \frac{2L}{\sigma^2_\text{r}}\Re\{\text{Tr}\{\mathbf{G}_j\mathbf{R}_\text{w}\mathbf{G}_i^H\}\},
\end{align}\end{subequations}
where we have used the fact that $\mathbb{E}\{\mathbf{SS}^H\} = L\mathbf{I}_{N_\text{t}+K}$ and we define $\mathbf{R}_\text{w}\triangleq\mathbf{WW}^H$. The other elements in $\mathbf{F}$ can be obtained based on its symmetry. The diagonal elements of the CRB matrix $\mathbf{C}=\mathbf{F}^{-1}$  represent the CRB for the elements of $\bm{\xi}$. To focus on the angle estimation performance, we partition $\mathbf{F}$ into $2\times 2$ blocks as 
\be\label{eq:FIM partition}
\mathbf{F} = \left[\begin{array}{cc}\mathbf{F}_1 & \mathbf{F}_2\\ \mathbf{F}^T_2 & \mathbf{F}_4\end{array}\right],
\ee
where the dimensions of $\mathbf{F}_1$, $\mathbf{F}_2$ and $\mathbf{F}_4$ are $2\times 2$, $2\times 2T$, and $2T\times 2T$, respectively. The submatrices $\mathbf{F}_1$, $\mathbf{F}_2$ and $\mathbf{F}_4$ are functions of $\mathbf{R}_\text{w}$ according to the definitions of their elements in (\ref{eq:F elements}). Using the properties of the Schur complement, the CRB for estimating $\theta_0$ and $\Delta$ is expressed as 
$\text{Tr}\{[\mathbf{F}_1-\mathbf{F}_2^T\mathbf{F}_4^{-1}\mathbf{F}_2]^{-1}\}$.

In this paper, we consider the ISAC problem of optimizing $\mathbf{W}$ to minimize the CRB for estimating $\theta_0$ and $\Delta$, while ensuring that the communication SINR requirements and transmit power constraint are met. The transmit beamforming optimization problem is formulated as 
\begin{subequations}\label{eq:original problem}\begin{align}
&\underset{\mathbf{W}}\min~~\text{Tr}\big\{\big[\mathbf{F}_1-\mathbf{F}_2^T\mathbf{F}_4^{-1}\mathbf{F}_2]^{-1}\big\}\label{eq:original problem obj}\\
&~~\text{s.t.}~~\frac{|\mathbf{h}_k^T\mathbf{w}_k|^2}
{\sum_{j\neq k}^{N_\text{t}+K}|\mathbf{h}_k^T\mathbf{w}_j|^2+\sigma_k^2}\geq \Gamma_k,~\forall k, \label{eq:op c1}\\
&\qquad~~\|\mathbf{W}\|_F^2 \leq P.
\end{align}\end{subequations}
This problem is highly non-convex due to the complicated objective function and fractional constraints. In the next section we propose an efficient algorithm to address these difficulties.

\section{CRB-Orientated Beamforming Design}

We first introduce an auxiliary variable $\mathbf{J}\in\mathbb{C}^{2\times 2}$, $\mathbf{J}\succeq\mathbf{0}$, to convert the objective function (\ref{eq:original problem}a) into
\begin{subequations}\begin{align}
&\underset{\mathbf{J},\mathbf{W}}\min~~\text{Tr}\{\mathbf{J}^{-1}\}\\
&~\text{s.t.}~~~\mathbf{F}_1-\mathbf{F}_2^T\mathbf{F}_4^{-1}\mathbf{F}_2\succeq \mathbf{J}.\label{eq:bf schur}
\end{align}\end{subequations}
Again using the Schur complement, (\ref{eq:bf schur}) can be further rewritten as
\be
\left[\begin{matrix} \mathbf{F}_1-\mathbf{J}& \mathbf{F}_2\\ \mathbf{F}_2^T& \mathbf{F}_4\end{matrix}\right]\succeq\mathbf{0}.
\ee
Then, considering that the expressions for the Fisher information matrix, communication SINR, and transmit power are all functions of $\mathbf{W}$ through the matrices $\mathbf{R}_\text{w}\triangleq\mathbf{WW}^H$ and $\mathbf{R}_k\triangleq\mathbf{w}_k\mathbf{w}_k^H,~\forall k$, we transform the optimization problem to
\begin{subequations}\label{eq:bf SDP}\begin{align}
&\underset{\mathbf{J},\mathbf{R}_\text{w},\mathbf{R}_k,\forall k}\min
~~\text{Tr}\{\mathbf{J}^{-1}\}\\
&\qquad\text{s.t.}\quad \left[\begin{matrix} \mathbf{F}_1-\mathbf{J}& \mathbf{F}_2\\ \mathbf{F}_2^T& \mathbf{F}_4\end{matrix}\right]\succeq\mathbf{0},\\
&\qquad\qquad~(1+\Gamma_k^{-1})\mathbf{h}_k^T\mathbf{R}_k\mathbf{h}_k^* -\mathbf{h}_k^T\mathbf{R}_\text{w}\mathbf{h}_k^*\geq \sigma_k^2,~\forall k, \\
&\qquad\qquad~\text{Tr}\{\mathbf{R}_\text{w}\} \leq P,\\
&\qquad\qquad~\mathbf{J}\in\mathcal{S}_2^+,~\mathbf{R}_\text{w}\in\mathcal{S}_{N_\text{t}}^+,~\mathbf{R}_k\in\mathcal{S}_{N_\text{t}}^+,~\forall k,\\
&\qquad\qquad~\mathbf{R}_\text{w}-\sum\nolimits_{k=1}^K\mathbf{R}_k\in\mathcal{S}_{N_\text{t}}^+,\\
&\qquad\qquad~\text{Rank}\{\mathbf{R}_k\}=1,~\forall k. \label{eq:rank-one constraint}
\end{align}\end{subequations}

To solve this problem, we use the classical semi-definite relaxation (SDR) technique, temporarily remove the rank-one constraint (\ref{eq:rank-one constraint}), and convert it to a semi-definite programming (SDP) problem that can be readily solved using various standard algorithms. As shown in \cite{Liu TSP 2022}, after obtaining the optimal solutions $\widetilde{\mathbf{R}}_\text{w}$ and $\widetilde{\mathbf{R}}_k$ to the relaxed problem, we can compute the optimal solutions $\mathbf{R}_\text{w}^\star$ and  $\mathbf{R}_k^\star$ to problem (\ref{eq:bf SDP}) and the corresponding beamformer $\mathbf{w}_k^\star$ as:
\begin{subequations}\label{eq:Rstar}\begin{align}
\mathbf{R}_\text{w}^\star &= \widetilde{\mathbf{R}}_\text{w}, \\
\mathbf{w}_k^\star &= (\mathbf{h}_k^T\widetilde{\mathbf{R}}_k\mathbf{h}_k^*)^{-1/2}\widetilde{\mathbf{R}}_k\mathbf{h}_k^*,~\forall k,\label{eq:initial wk}\\
\mathbf{R}_k^\star &=  \mathbf{w}_k^\star(\mathbf{w}_k^\star)^H.
\end{align}\end{subequations}
Recalling that $\mathbf{R}_\text{w} = \mathbf{WW}^H=\mathbf{W}_\text{c}\mathbf{W}_\text{c}^H+\mathbf{W}_\text{r}\mathbf{W}_\text{r}^H$, the radar beamformer $\mathbf{W}_\text{r}$ should satisfy
\be\label{eq:initial Wr}
\mathbf{W}_\text{r}\mathbf{W}_\text{r}^H = \mathbf{R}_\text{w}^\star-\sum_{k=1}^K\mathbf{R}_k^\star,
\ee
from which we can calculate $\mathbf{W}_\text{r}^\star$ using a Cholesky or eigenvalue decomposition.

\section{CRB Optimization using UCM and DSM}

To demonstrate the advantages of the PSM and its associated CRB-oriented beamforming design, this section discusses CRB optimization using the previously studied UCM and the DSM approaches. A method for fairly comparing the different models is then provided.

\subsection{UCM-based CRB Optimization}

UCM treats the extended target response $\mathbf{G}$ as an unstructured matrix and directly estimates it. Thus the parameters of the UCM are the elements of $\mathbf{G}$ themselves, and the corresponding Fisher information matrix $\mathbf{G}$ is given by
\be
\mathbf{F}_{\mathbf{G}} = \frac{L}{\sigma^2_\text{r}}(\mathbf{R}_\text{w}\otimes \mathbf{I}_{N_\text{r}}),
\ee
from which the CRB for estimating $\mathbf{G}$ can be calculated as
\be
\mathbf{C}_{\mathbf{G}} = \mathbf{F}_\mathbf{G}^{-1}.
\ee
Thus, the UCM-based beamforming design attempts to minimize $\text{Tr}\{\mathbf{R}_\text{w}^{-1}\}$ while satisfying the communication SINR and transmit power constraints \cite{Liu TSP 2022}. An algorithm similar to that in Section IV can be employed to solve the resulting optimization problem. However, UCM fails to exploit knowledge of the radar array responses, leading to an omnidirectional beampattern. 
Moreover, the number of parameters to be estimated using UCM is relatively large, especially for large arrays, making UCM less efficient for practical applications.

\subsection{DSM-based CRB Optimization}
DSM treats each scatterer as an independent source of reflection from all others, requiring the estimation of the following set of parameters: 
$\widetilde{\bm{\xi}}\triangleq[\theta_1,\theta_2,\ldots,\theta_T,\Re\{\bm{\alpha}^T\},\Im\{\bm{\alpha}^T\}]^T$. The corresponding Fisher information matrix $\mathbf{F}_{\widetilde{\bm{\xi}}}\in\mathbb{R}^{3T\times 3T}$ can be calculated using the expressions previously derived in~(\ref{eq:G}) 
and~(\ref{eq:vector signal}). Then, following the same procedure as in (\ref{eq:FIM partition}), the Fisher information matrix is partitioned into  $T\times T$, $T\times 2T$, $2T\times T$, and $2T\times 2T$ sub-matrices, and the CRB for angle estimation can be obtained as the objective function to be minimized. Although compared to UCM, DSM reduces the parameter dimensionality to $3T$, it is still tends to overparameterize the problem since it focuses on detailed local information for each scatterer without considering their relationship to each other across the extended target. This overparameterization can reduce the parameter estimation accuracy, and it introduces additional complexity to both the beamforming design and the estimation algorithm.

\subsection{Jacobian-based Transformation for estimating $\bm{\xi}$}
To ensure a fair comparison the angle estimation performance achievable by the three models, the CRB results from the UCM and DSM-based optimizations must be converted to find the performance bound for estimating $\bm{\xi}$ using their particular parameterization. This involves transforming the CRB for the UCM and DSM parameters (i.e., the response matrix $\mathbf{G}$ or the independent angles $\{\theta_1,\theta_2,\ldots,\theta_T\}$) into the corresponding CRB for the parameters in $\bm{\xi}$. To do so, we define the Jacobian matrices $\mathbf{J}_{\mathbf{G}}(\bm{\xi})\in\mathbb{C}^{N_\text{t}N_\text{r}\times (2T+2)}$ and $\mathbf{J}_{\widetilde{\bm{\xi}}}(\bm{\xi})\in\mathbb{C}^{3T\times(2T+2)}$ as 
\begin{subequations}\begin{align}
\mathbf{J}_{\mathbf{G}}(\bm{\xi}) &\triangleq \Big[\frac{\partial\text{vec}\{\mathbf{G}\}}{\partial\theta_0} \;\;\; \frac{\partial\text{vec}\{\mathbf{G}\}}{\partial\Delta} \;\;\; \frac{\partial\text{vec}\{\mathbf{G}\}}{\partial\Re\{\bm{\alpha}\}^T} \;\;\; \frac{\partial\text{vec}\{\mathbf{G}\}}{\partial\Im\{\bm{\alpha}\}^T}\Big]\\
   \mathbf{J}_{\widetilde{\bm{\xi}}}(\bm{\xi}) &\triangleq \Big[\frac{\partial\widetilde{\bm{\xi}}}{\partial\theta_0} \;\;\; \frac{\partial\widetilde{\bm{\xi}}}{\partial\Delta} \;\;\; \frac{\partial\widetilde{\bm{\xi}}}{\partial\Re\{\bm{\alpha}\}^T} \;\;\; \frac{\partial\widetilde{\bm{\xi}}}{\partial\Im\{\bm{\alpha}\}^T}\Big],
\end{align}\end{subequations}
where the partial derivatives can be calculated based on the relationships in~(\ref{eq:G}), (\ref{eq:G derivatives}) and (\ref{eq:thetat}).
Then, the Fisher information matrix for estimating $\bm{\xi}$ using UCM and DSM can be respectively calculated as 
\begin{subequations}
\begin{align}
\mathbf{F}_{\bm{\xi}} &= \mathbf{J}^H_{\mathbf{G}}(\bm{\xi})\mathbf{F}_\mathbf{G}\mathbf{J}_{\mathbf{G}}(\bm{\xi}),\\
\mathbf{F}_{\bm{\xi}} &= \mathbf{J}^H_{\widetilde{\bm{\xi}}}(\bm{\xi})\mathbf{F}_{\widetilde{\bm{\xi}}}\mathbf{J}_{\widetilde{\bm{\xi}}}(\bm{\xi}).
\end{align}\end{subequations}

\section{Simulations}
In this section, we present simulation results to demonstrate the advantages of the proposed CRB optimization using PSM for extended targets in ISAC systems. Unless otherwise specified, we assume $N_\text{t} = 20$, $N_\text{r}=60$, $K=4$, $P=30$dBm, $L = 30$, $\Gamma_k=\Gamma=10$dB, $\forall k$, $\sigma_k^2=\sigma_\text{r}^2 = 0$dBm, $\forall k$. The extended target is positioned at a central angle of $\theta_0 = 30^\circ$ with an angular spread of $\Delta = 6^\circ$, consisting of $T=4$ scatters, each with an RCS of $\alpha_t = 0.001,~\forall t$. For the communication channels, a Rician fading model is adopted with a Rician factor of 10. 


\begin{figure}[!t]
    \centering
    \includegraphics[width=0.85\linewidth]{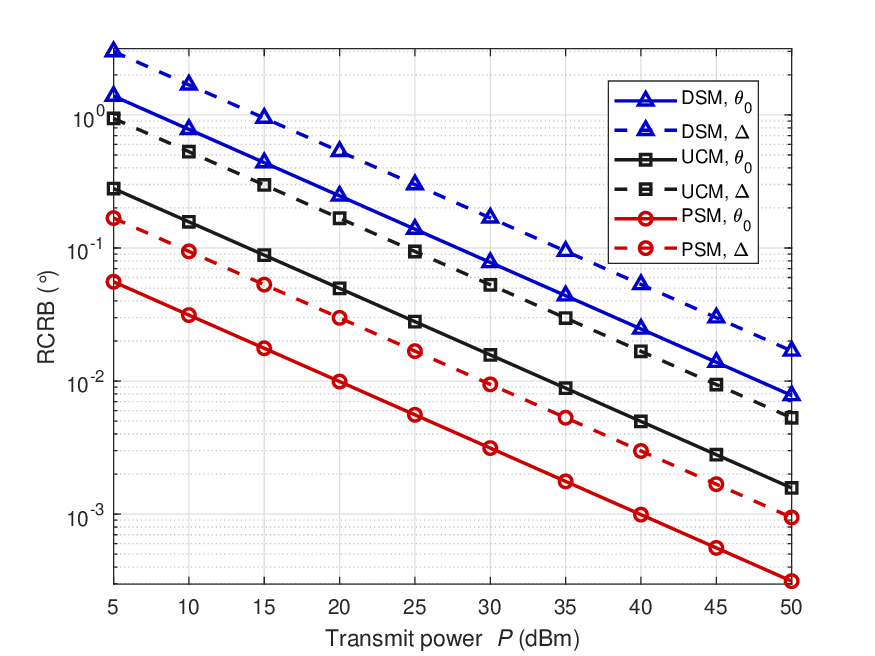}
    \caption{RCRB versus transmit power $P$.}\vspace{-0.2 cm}
    \label{fig:RCRB_power}
\end{figure}

We first show the root-CRB (RCRB) versus the transmit power in Fig.~\ref{fig:RCRB_power}, where solid lines represent the RCRB for $\theta_0$ and dashed lines represent the RCRB for $\Delta$. It can be observed that the CRB for $\Delta$ is consistently higher than that for $\theta_0$, highlighting the increased difficulty associated with estimating the angular spread. We also clearly observe that the optimization based on the proposed PSM approach yields the lowest CRB, demonstrating its effectiveness in capturing the essential parameters for extended targets. In contrast, the two-step approaches based on UCM and DSM yield higher CRBs due to their inherent limitations in exploiting the structure of the extended target's channel response.

\begin{figure}[!t]
    \centering
    \includegraphics[width=0.85\linewidth]{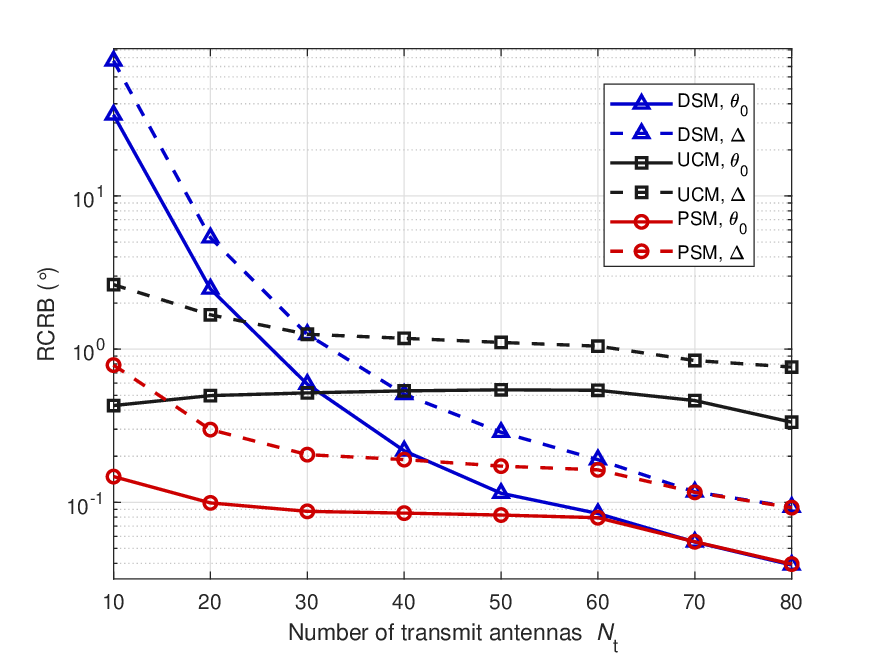}
    \caption{RCRB versus number of transmit antennas $N_\text{t}$.}\vspace{-0.2 cm}
    \label{fig:RCRB_Nt}
\end{figure}

\begin{figure}[!t]
    \centering
    \includegraphics[width=0.85\linewidth]{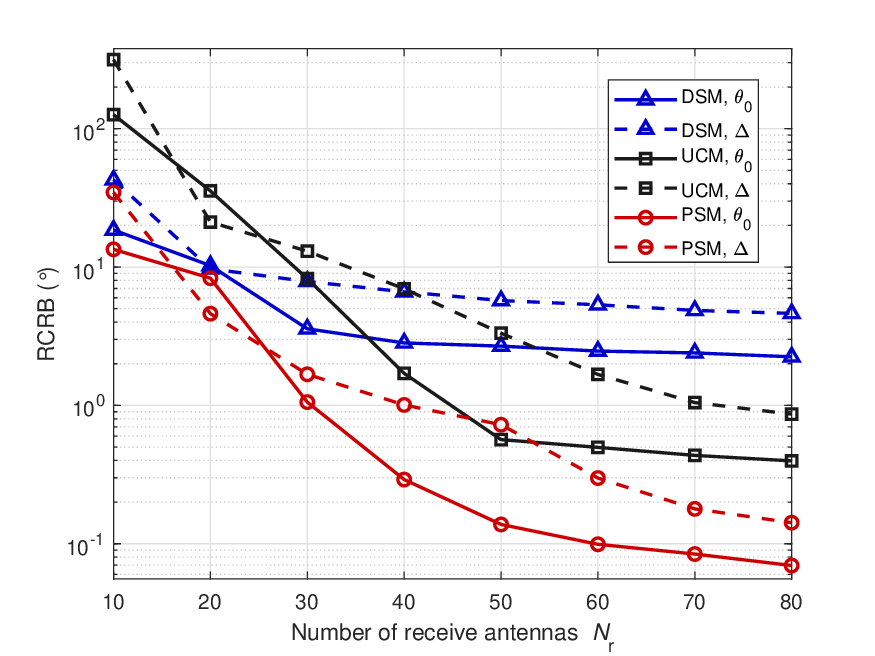}
    \caption{RCRB versus number of receive antennas $N_\text{r}$.}\vspace{-0.2 cm}
    \label{fig:RCRB_Nr}
\end{figure}

Fig.~\ref{fig:RCRB_Nt} illustrates the RCRB as a function of the number of transmit antennas. As expected, increasing the number of transmit antennas enhances beamforming gains, thereby improving parameter estimation performance across all approaches. This improvement becomes particularly pronounced when the number of transmit antennas exceeds the number of receive antennas, i.e., $N_\text{t} > 60$, thanks to enhanced spatial diversity, which offers additional flexibility for beamforming and parameter estimation. However, the DSM approach benefits the most from a larger transmit antenna array. Since it relies heavily on resolving the angles of individual scatterers, the increased number of antennas enhances beamforming resolution, enabling more accurate angular estimation and reducing overall estimation error. Despite this improvement, the DSM approach still lags slightly behind the PSM approach due to its over-parameterization. In contrast, the UCM approach shows minimal improvement with additional transmit antennas, as it treats the target response as an unstructured matrix, failing to leverage angular information for refined beamforming. 
Notably, the PSM approach can provide superior performance even with a small number of transmit antennas. By focusing on a more parsimonious representation of the target, the PSM efficiently captures the characteristics of the extended target without the need for high beamforming resolution. This highlights the advantages of the PSM approach in scenarios with limited spatial resources.

The RCRB versus the number of receive antennas is presented in Fig.~\ref{fig:RCRB_Nr}. The CRB decreases with more receive antennas thanks to improved spatial resolution and diversity. However, this improvement tends to plateau once the available spatial information is fully exploited. It is noted that the PSM approach maintains satisfactory performance even with a moderate number of antennas, again highlighting its efficiency and practical advantages without relying on large-scale arrays.

\section{Conclusions}
In this paper, we introduced a parametric scattering model (PSM) to efficiently capture the angular characteristics of an extended target,  along with the CRB corresponding to estimates of these parameters. The CRB was minimized under transmit power and communication SINR constraints. Simulation results confirmed that the proposed PSM approach outperforms traditional approaches based on other extended target models, especially in scenarios where the spatial resources are limited. Future work will explore extending the model to orthogonal frequency division multiplexing (OFDM) systems, where improved range resolution will extend the target visibility in the range domain, adding a new dimension to the estimation challenge.


\begin{thebibliography}{99}
\bibitem{LiuFan JSAC 2022} F. Liu, Y. Cui, C. Masouros, J. Xu, T. X. Han, Y. C. Eldar, and S. Buzzi, ``Integrated sensing and communications: Towards dual-functional wireless networks for 6G and beyond,'' \textit{IEEE J. Sel. Area Commun.}, vol. 40, no. 6, pp. 1728-1767, Jun. 2022.

\bibitem{Zhang ICST 2022} J. A. Zhang \textit{et al.}, ``Enabling joint communication and radar sensing in mobile networks — A survey,'' \textit{IEEE Commun. Surveys Tuts.}, vol. 24, no. 1, pp. 306-345, First Quart. 2022.

\bibitem{Rang JSTSP 2022} R. Liu, M. Li, Y. Liu, Q. Wu, and Q. Liu, ``Joint transmit waveform and passive beamforming design for RIS-aided DFRC systems,'' \textit{IEEE J. Sel. Topics Signal Process.}, vol. 16, no. 5, pp. 995-1010, Aug. 2022.

\bibitem{AK2024} A. Kaushik \textit{et al.}, ``Toward integrated sensing and communications for 6G: Key enabling technologies, standardization, and challenges,'' \textit{IEEE Commun. Standards Mag.}, vol. 8, no. 2, pp. 52-59, Jun. 2024.


\bibitem{Liu TSP 2022} F. Liu, Y.-F. Liu, A. Li, C. Masouros, and Y. C. Eldar, ``Cram{\'e}r-Rao bound optimization for joint radar-communication beamforming,'' \textit{IEEE Trans. Signal Process.}, vol. 70, pp. 240-253, Dec. 2021.

\bibitem{Hua TWC 2024} H. Hua, T. X. Han, and J. Xu, ``MIMO integrated sensing and communication: CRB-rate tradeoff,'' \textit{IEEE Trans. Wireless Commun.}, vol. 23, no. 4, pp. 2839-2854, Apr. 2024.

\bibitem{Song TWC 2024} X. Song, X. Qin, J. Xu, and R. Zhang, ``Cram{\'e}r-Rao bound minimization for IRS-enabled multiuser integrated sensing and communications,'' \textit{IEEE Trans. Wireless Commun.}, vol. 23, no. 8, pp. 9714-9729, Aug. 2024.


\bibitem{Karbasi TSP 2015} S. M. Karbasi, A. Aubry, A. De Maio, and M. H. Bastani, ``Robust transmit code and receive filter design for extended targets in clutter,'' \textit{IEEE Trans. Signal Process.}, vol. 63, no. 8, pp. 1965-1976, Apr. 2015. 

\bibitem{Yao TITS 2022} Y. Yao, H. Liu, P. Miao, and L. Wu, ``MIMO radar design for extended target detection in a spectrally crowded environment,'' \textit{IEEE Trans. Intell. Transp. Syst.}, vol. 23, no. 9, pp. 14389-14398, Sep. 2022.

\bibitem{Du TWC 2023} Z. Du, F. Liu, W. Yuan, C. Masouros, Z. Zhang, S. Xia, and G. Caire, ``Integrated sensing and communications for V2I networks: Dynamic predictive beamforming for extended vehicle targets,'' \textit{IEEE Trans. Wireless Commun.}, vol. 22, no. 6, pp. 3612-3627, Jun. 2023. 




\end{thebibliography}
\end{document}